# Is one-way modal conversion sufficient for optical isolation?

*Comment on "Nonreciprocal light propagation in a silicon photonic circuit", by Feng et al, Science 333, 729 (2011)*

Zongfu Yu[1], J. D. Joannopoulos[2], and Shanhui Fan[1*]

[1]Ginzton Laboratory, Department of Electrical Engineering, Stanford University, Stanford, CA 94305
[2]Department of Physics, Massachusetts Institute of Technology, Cambridge, MA 02139

[*]To whom the correspondence should be addressed. Email: shanhui@stanford.edu

**Abstract**

We show that the novel one-way photonic modal conversion effect, as demonstrated by Feng et al (Reports, 11 August, 2011, p. 729), is insufficient in itself to enable optical isolation, since the underlying dielectric structure possesses a symmetric scattering matrix. Moreover, one cannot construct an optical isolator, by enclosing this structure with a system containing any combination of components or signal processing elements, as long as the overall system is linear, and is described by a static scalar dielectric function. Instead, to achieve complete optical isolation, one would need to augment the present design so that the final system is described by a *non-symmetric* scattering matrix.

In [1], Feng et al considered a two-mode waveguide with a statically modulated dielectric constant (Fig. 1, left panels), and demonstrated a novel one-way modal conversion effect: The odd mode is strongly excited, when an even mode is incident along the backward direction from the right end of the waveguide (Fig. 1b). Whereas the odd mode is not excited, when the same even mode is incident along the forward direction from the left end (Fig. 1a).

The structure in [1] is certainly of fundamental and practical interest. For example, one would naturally like to use such a structure to achieve non-magnetic optical isolation. However, is the observed effect of one-way photonic modal conversion sufficient for constructing an optical isolator?

Unfortunately, the answer to the above question is No. One cannot construct an optical isolator this way.

It is well known that any linear system described by a *scalar* dielectric function $\epsilon(r)$, including the structure of [1], is constrained by the reciprocity theorem [2]. Such a system is *reciprocal*, in the sense that its scattering matrix is symmetric [2]. Very importantly, the reciprocity theorem applies even when $\epsilon(r)$ is *complex*, i.e. even when the system has gain or loss.

The system in [1] indeed has a symmetric scattering matrix. In Figs. 1a and 1b, (which are equivalent to Fig. S1 in the Supplementary Information of [1]), one injects the even mode along either forward or backward directions. Notice that, as light propagates, the photon fluxes in the even mode (red curves in Figs. 1a and 1b) are the same for both directions, as expected from the reciprocity theorem.

It is instructive to contrast the static system in [1], to the dynamic system in [3], which applied a time-dependent dielectric modulation $\Delta\epsilon = \delta\cos(kx - \Omega t)$ to the waveguide. In the time-dependent system, when the even mode is injected, the photon fluxes in the even mode depend on the directions of light injection (red curves in Figs. 1d and 1e), a clear indication of non-reciprocity.

Why doesn't the structure in [1] function as an optical isolator? Certainly, in this structure, the even mode injected from the left is strongly attenuated without exciting the odd mode (Fig. 1a), whereas the even mode injected from the right excites the odd mode and hence has a substantial power transmission coefficient $T$ (Fig. 1b). Thus, there appears to be a contrast between the two propagation directions. However, one should not confuse such a contrast with what is required of an optical isolator. An optical isolator needs to completely suppress any back reflection. But in this structure, at the left end, light reflected back into the odd mode will necessarily pass through the structure with the same high power transmission coefficient $T$ (Fig. 1c, red curve). Therefore, the structure cannot provide *complete* optical isolation: It provides isolation only when the reflected light is in the even mode. It does *not* provide isolation for *every* possible reflection.

More generally, one cannot construct an optical isolator, with *any* structure having a *symmetric* scattering matrix. And thus, one cannot construct an optical isolator in any linear system described by a static and scalar dielectric function, no matter how complicated the geometry of the design. As an illustration, consider a general two-port system (Fig. 2), where two single-mode waveguides are coupled to a general structure. This general structure may enclose the structure in [1], together with any other components or signal processing elements. If this general system is described by a scalar dielectric function, its scattering matrix $S$, defined as

$$\begin{pmatrix} b_1 \\ b_2 \end{pmatrix} = S \begin{pmatrix} a_1 \\ a_2 \end{pmatrix} \equiv \begin{pmatrix} r_1 & t_{12} \\ t_{21} & r_2 \end{pmatrix} \begin{pmatrix} a_1 \\ a_2 \end{pmatrix}$$

must be symmetric [2]. As a result, we have

$$t_{12} = t_{21}$$

and hence the transmission coefficients from either direction must be equal in both amplitude and phase. This general two-port system therefore cannot function as an optical isolator.

To summarize, [1] demonstrated a novel optical structure. However, the structure in [1] has a symmetric scattering matrix, which fundamentally differentiates it from an optical isolator structure. Moreover, one cannot construct an optical isolator, by simply enclosing the structure in [1] with a system containing any combinations of components or signal processing elements, as long as the overall system is linear, and is described by a static scalar dielectric function. In order to use the design in [1] to enable complete optical isolation, one would need to augment it so that the scattering matrix of the final system is non-symmetric.

**Figure Captions**

**Figure 1:** Photon flux (N) in the even (red) and odd (blue) modes in a two-mode waveguide parallel to the *x*-axis, as a function of propagation distance, as calculated using coupled mode theory of [1] and [3]. The propagation distance is normalized with respect to a length scale related to the modal coupling constant. The insets show the direction and mode shape of the input and output light. The gray region is the waveguide. The dark region is modulated. (**a-c**) Reciprocal structure. (**d-f**) Non-reciprocal structure.

**Figure 2:** A general two-port system consists of two single-mode waveguides (gray) couple to a general structure. $a_{1,2}$ and $b_{1,2}$ are the input and output wave amplitudes, respectively.

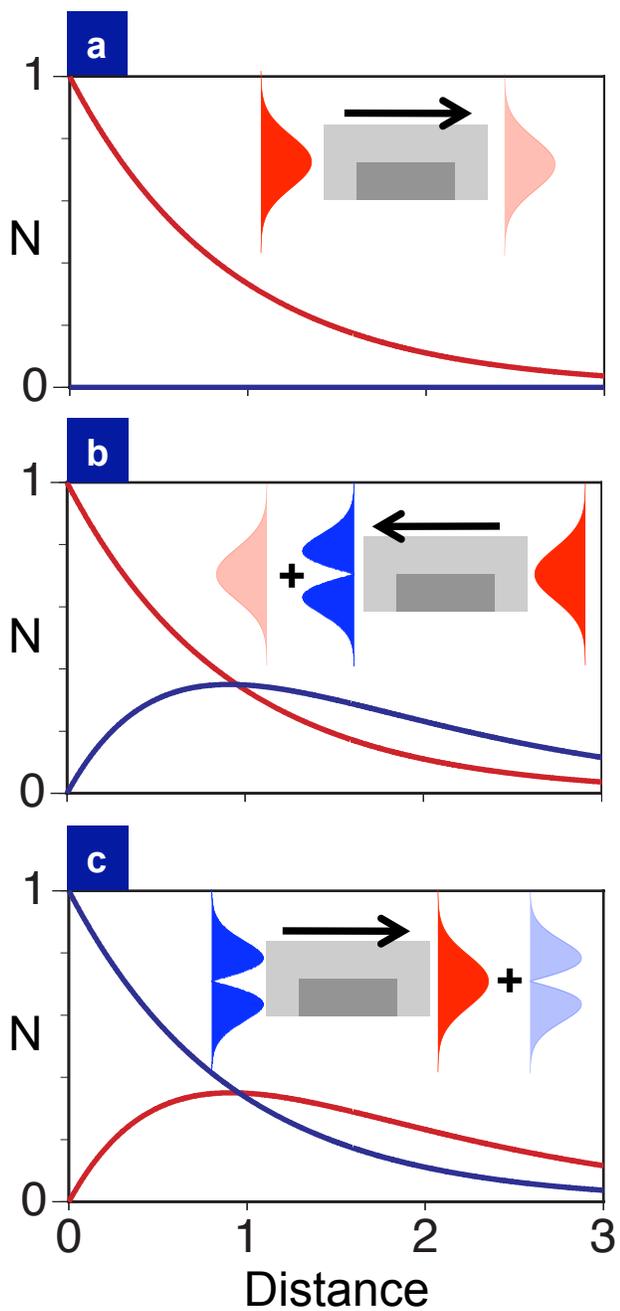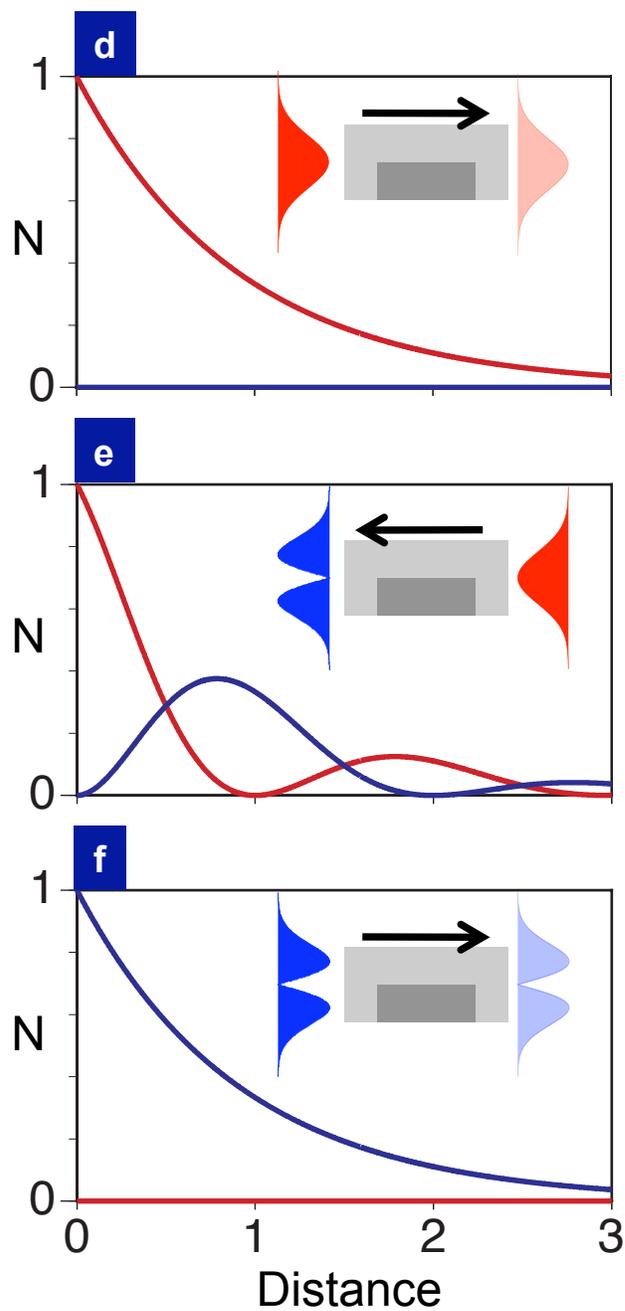

**Figure 1**

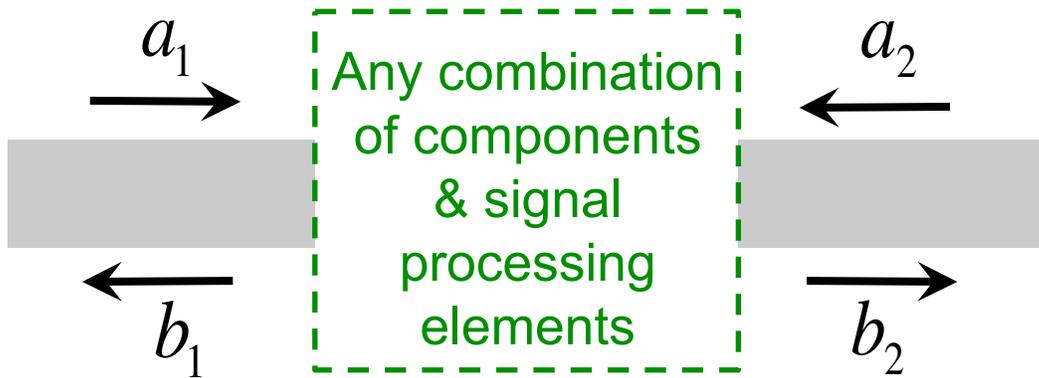

**Figure 2**